# The Rôle of $\gamma_5$ in Dimensional Regularization

Dirk Kreimer*

Dept. of Physics

Univ. of Tasmania

G.P.O. Box 252C

Hobart 7001

Australia

December 1993

**Abstract**

We discuss the consistency of a new $\gamma_5$-scheme with renormalization. In particular we study the power-counting behaviour of multiloop graphs to prove its consistency. As a side effect we obtain a short proof of the Adler-Bardeen theorem. Further we show that this $\gamma_5$-scheme does not modify the BRST identities at any loop orders in contrast to BM type schemes.

## 1 Introduction

At the moment the situation in perturbation theory using dimensional regularization (DR) as a regulator and investigating problems involving $\gamma_5$ is somewhat weird. On the one hand, there is the proposal of abandoning the anticommuting $\gamma_5$ as in [1]. This scheme emphasizes the loss of full $D$-dimensional Lorentz invariance by introducing $D$-dimensional, 4-dimensional and $(D-4)$-dimensional objects. It restricts the range of indices in metric and Levi-Civita tensors. A similar approach, based on physical demands of axial amplitudes was obtained in [2]. It avoids the difficult notion of a tensor with indices of restricted range by considering antisymmetrized products of $\gamma$-matrices instead of $\gamma_5$, as it is motivated for example by the use of the Levi-Civita tensor contracted with three $\gamma$-matrices for the chiral current. These schemes, in the following collectively denoted as BM schemes, give a well-defined regulator for arbitrary applications of $\gamma_5$. They are well-defined in the sense that they respect the structure of the loop expansion of the full theory, which means that the renormalization program can be carried out [1].

Unfortunately these schemes violate the BRST Identities and one thus has to install them by hand order by order [3]. A more systematic approach seems likely but is still not available.

On the other hand, in [4] a scheme was proposed which suggests a different treatment of the $\gamma_5$-problem. There is overwhelming evidence [5] that the use of an anticommuting $\gamma_5$ is legitimate in anomaly free theories. Preserving an anticommuting $\gamma_5$, the proposed scheme is based on a replacement of the usual trace in four dimensions by a functional $\mathcal{T}$race, a linear functional which agrees with the trace only in $D=4$ dimensions. The reader will find all necessary definitions and properties in an appendix.

*email: kreimer@physvax.phys.utas.edu.au

It is a feature of this new scheme that one can establish a one to one correspondence between the $\gamma_5$-problem and the appearance of anomalies. In this paper we will prove that this scheme is well defined in the above sense, that is it is compatible with the renormalization program. Further we will prove that no violation of BRST identities can appear to any loop orders, and we will show that the Adler-Bardeen theorem arises quite naturally in the context of this scheme.

The organization of the paper is as follows. First we give some general arguments excluding reading prescriptions which would immediately lead to wrong results. We will insist on charge conjugation properties (Furry's theorem) and Bose symmetry to restrict the number of allowed reading prescriptions. We will end up with a final reading prescription which is applicable in the most general circumstances including applications of DR in infrared problems or as a regulator in non-renormalizable theories.

Then we show the promised conservation of BRST identities. The crucial point is the investigation of the BRST identity for the fermionic vertex, as its radiative corrections are possible candidates for the generation of an non-anticommuting $\gamma_5$ in higher loop orders, even in this scheme. The Adler-Bardeen theorem then follows from a screening mechanism, where, de facto, Zimmermann's forest formula protects the axial ($A$) vertex inside a forest from being anomalous.

Some more general remarks on the rôle of $\gamma_5$ will finish this paper.

We have collected all relevant formulas and derivations of the main algebraic properties of the $\mathcal{T}$race functional in an appendix, most of them the reader will also find in [6]. In abandoning cyclicity we are forced to choose reading points. Some necessary notation for this is collected in an appendix too, as well as our notation for the forest formula.

## 2   Defining the Scheme

Given an arbitrary string of $\gamma$-matrices the question arises where one should start reading it when one applies a non-cyclic functional to it. One of the necessary conditions is that one should be able to obtain unique results for a renormalizable theory. This includes the situation that subdivergences of a fermion loop are subtracted in a consistent manner. In the following we first discuss these subdiagrams of a fermion loop and extract a condition on reading prescriptions from these considerations. We then investigate how some basic facts resulting from the study of one-loop anomalous graphs lead to further constraints for reading prescriptions. In doing so we abandon all reading prescriptions which could be only defined with respect to a specified topology of a graph, and we insist on Bose symmetry and charge conjugation properties as they are expressed in Furry's theorem for fermion loops.

In this way we obtain as the final result of this section a unique reading prescription which has to be considered as part of the definition of our scheme.

Our starting point to find sensible reading prescriptions is to consider a closed fermion loop in an anomaly free theory. An abnormal amplitude (by this we mean an amplitude involving a $\gamma_5$-odd fermion loop) then has a vanishing overall degree of divergence. We allow for arbitrary subloops. We want to find a sufficient condition to have unique results for the $\mathcal{T}$race functional under these circumstances. To this end consider the associated forests of its subdivergences. Every forest has its maximal forest, and forests never overlap. So in general the fermion loop will look like the following example in Fig.(1).

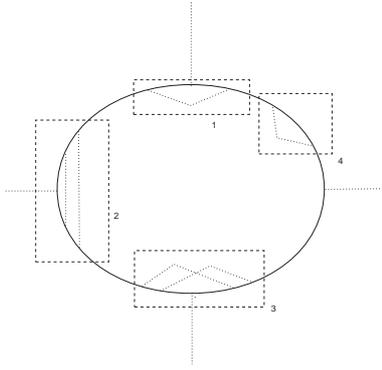

Fig.(1) A fermion loop with subdivergences and forest structure including $k = 4$ maximal forests.

A chosen reading point will be either outside or inside a maximal forest. A precise definition of the notion of a reading point inside or outside a maximal forest is given in the appendix. Every forest corresponds to a subdivergence and it is a crucial ingredient of the renormalization program that this subdivergence corresponds to a unique Laurent series in the cut-off parameter ($(D - 4)$ in DR). We can calculate the subdivergences separately and the resulting counterterms must cancel the subdivergences of Fig.(1).

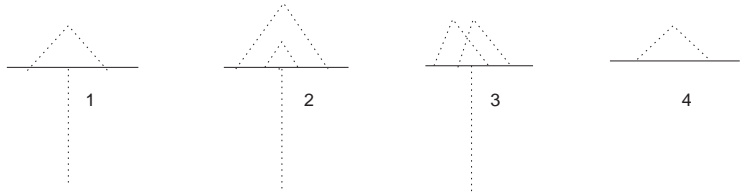

Fig.(2) The subdivergences of Fig.(1).

To be able to calculate Fig.(1) we have to choose a reading point. Assume we have chosen one not located inside a maximal forest, then $\mathcal{T}$race would result in an expression of the form

$$\mathcal{T}\mathrm{r}(\gamma_5 \not{p}_1 \Gamma_1 \not{p}_2 \ldots \not{p}_k \Gamma_k),$$

where the $\Gamma_i$ in the above expression correspond to the strings of $\gamma$-matrices associated to the $k$ maximal forests of the fermion loop and the $p_i$ correspond to the propagators carrying the charge flow of the fermion loop between these maximal forests. The $\Gamma_i$ correspond on the other hand to the subdivergences generating the relevant counterterms, as in Fig.(2) above for our example.

We see that the addition of appropriate counterterms will cancel these subdivergences. As there are no overall divergences by assumption, adding all counterterms gives a finite result, $\overline{\mathcal{R}(G)}$ = finite. Now the difference of two reading prescriptions (each one based on reading points starting outside maximal forests) is an order $(D - 4)$ operator, vanishing when multiplied by a finite expression. So all these reading prescriptions are equivalent.

If we were to choose a reading point located inside a maximal forest, we would destroy the mutual cancellation between subdivergences and counterterms, which can be explicitly checked already at the two-loop level. We would have a $\mathcal{T}$race of the following form

$$\mathcal{T}\mathrm{r}(\Gamma_{k,f} \gamma_5 \not{p}_1 \Gamma_1 \ldots \Gamma_{k,i}),$$
$$\Gamma_k = \Gamma_{k,i} \Gamma_{k,f},$$

where the splitting of $\Gamma_k$ into pieces $\Gamma_{k,i}$ and $\Gamma_{k,f}$ indicates the modification of the subdivergence. We cannot use cyclicity to connect these open lines. So we would need a counterterm which differs from the one generated by the unbroken $\Gamma_k$. The following Fig.(3) gives an example how a two-loop subdivergence, cut open by an illegitimate reading prescription, fails to be compensated by the standard counterterms.

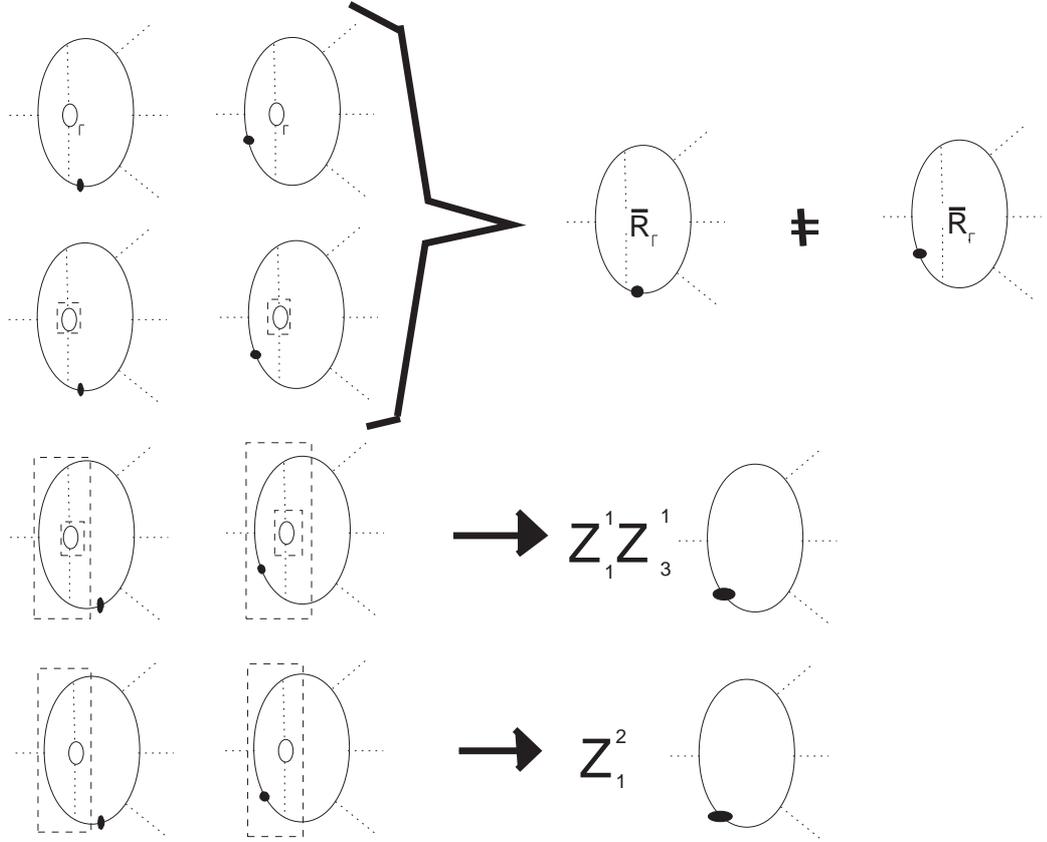

Fig.(3) A legal and an illegal reading prescription.

The first column of graphs on the left represents all forests (assuming vanishing overall degree of divergence). It has an allowed reading point outside the maximal forest. The second column has an illegal reading point. $\Gamma$ denotes the loop at the internal boson line. Assuming that all subdivergences give rise to the standard counterterms as they should, both columns give rise to similar counterterms for the last two graphs. They involve a one-loop Z-factor for the boson self-energy $Z_3^1$, a one-loop Z-factor for the vertex correction $Z_1^1$ and a two-loop Z-factor $Z_1^2$. In both columns the first two graphs can be summarized by replacing $\Gamma$ by $\overline{\mathcal{R}(\Gamma)}$. The two columns disagree. Calculating the fermion loop with a bare internal boson line gives a result $c_0 \frac{1}{D-4} + c_1$ say for the left column. For the right we would obtain $c_0 \frac{1}{D-4} + d_1$, with $c_1 \neq d_1$. So the finite parts disagree. But by dressing the internal boson line with $\overline{\mathcal{R}(\Gamma)}$ the leading divergence is $\sim \frac{1}{(D-4)^2}$. The subtraction involved in $\overline{\mathcal{R}(\Gamma)}$ subtracts the non-local bits and pieces but does not kill the leading divergence in $(D-4)$. This is so because we have $\mathcal{C}(\Gamma_2) \neq 0$, where $\Gamma_2$ denotes the two-loop subdivergence which generates the maximal forest. So the disagreement between the two columns appears now in the non-leading but still divergent terms $\sim \frac{1}{D-4}$. As the sum of the four graphs in the left column is finite because the last two graphs give just the counterterms for the subdivergences generated by the first two graphs we see that the same counterterms can not render the first two graphs in the right column finite.

We learn that we have to base reading prescriptions on reading points located outside of maximal forests to respect the subtraction of subdivergences of a closed fermion loop, as it is governed by the forest formula. We have a unique result for fermion loops with vanishing overall degree of divergence. We do not continue discussing renormalizable theories here. Up to now we have restricted ourselves to the study of amplitudes of vanishing overall degree of divergence. As we want to obtain a unique reading prescription we continue to find further restrictions for reading prescriptions. After we have done so and thus completed the definition of our scheme we are prepared for the more difficult problem of fermion loops being subloops in other graphs, thus suffering from overall divergences. We will attack this problem in the next section when we study BRST identities.

Let us consider the anomaly. In [4] it was shown how to extract the anomaly as the typically non-cyclic output of the $\gamma_5$-scheme considered here. One of the lessons to be learned from this is that we are not allowed to start $\mathcal{T}$race reading at different points in graphs contributing to the same process. In [6] an example was given how one could violate gauge invariance by erroneously doing so for the SM. Another lesson learnt from the study of the anomaly is that we should and can maintain Bose symmetry. Considering the $AAA$ anomaly there are six possibilities to connect the three vertices of the amputated Green function to three exterior lines. Three of these possibilities simply correspond to cyclic permutations and would be identical for a cyclic trace functional. They would cancel a factor 3 in $\frac{1}{3!}$, arising from the Taylor expansion of the interaction part of the generating functional for the loop expansion. For the noncyclic $\mathcal{T}$race we then have $\frac{1}{3} \times (3$ reading possibilities) [6]. In general we would have $\frac{1}{n} \times (\sum n$ reading points $)$ if the permutations generated by Bose symmetry contain the cyclic permutation group of order $n$ as a subgroup, see Fig.(4). The correct result $AAA = \frac{1}{3}AVV$ is generated by this constraint on reading prescriptions. Also the physical constraint of vector current conservation was used for the $AVV$ anomaly, which nonetheless, as we will see shortly, is only a constraint at the one-loop level.

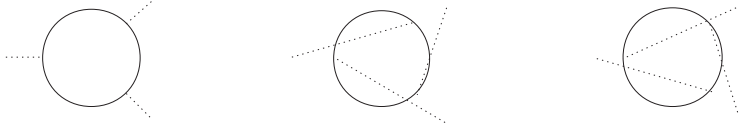

Fig.(4) The cyclic permutation subgroup of all Bose permutations for a fermion loop.

Up to now we have excluded the interior of maximal forests as possible reading points and we have established Bose symmetry for a chosen reading prescription. But what then are possible reading points and prescriptions?

As explained in the appendix there are type $i_+, i_I, i_{II}$ and type $III$ reading points. They correspond to the first vertex of a maximal forest of an exterior vertex (type $i_+$), to the outgoing propagator of this forest (type $i_I$), and to the propagator entering this forest (type $i_{II}$) respectively. The rest are type $III$ reading points which are located between maximal forests of self energy corrections of propagators carrying the charge flow of the fermion loop:

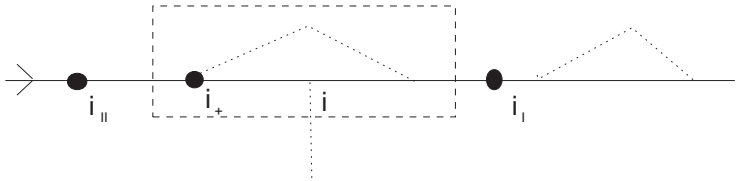

Fig.(5a) The different types of reading points. Here and in all following figures an arrow at the fermion line indicates the momentum flow, not the charge flow.

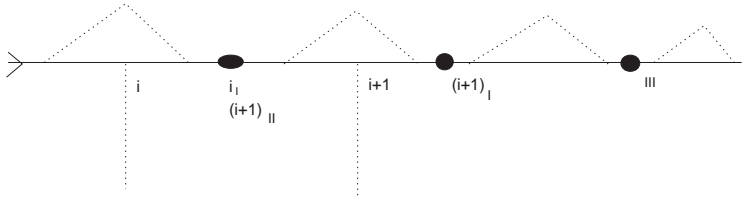

Fig.(5b) Propagators as reading points. Note that a propagator can be of type (I) and (II) similarly.

We do not consider exterior scalar or pseudo scalar vertices here as they correspond to type $III$ reading points, as explained in the appendix.

Further, we exclude type $III$ reading points from reading prescriptions as they are not identifyable in a unique manner, according to the various ways of dressing propagators with self-energies.

Studying the anomaly it was found in [4, 6] that type $i_I$ reading points are equivalent to reading points of type $i_+$. This resulted from the fact that a violation of a Ward identity is only possible at a given vertex if the momentum flow is cut open at this vertex.

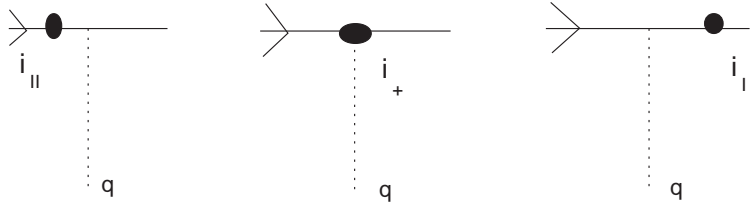

Fig.(6) The momentum flow at a vertex. BRST identities for this vertex can only be violated if the momentum flow is cut open, otherwise the substitution $\slashed{q} \to (\slashed{k} + \slashed{q}) - \slashed{k}$ would cancel a propagator to the left resp. right and thus establish the BRST identity for this vertex. The type $i_I$ and the type $i_+$ reading points give the same result.

In the same way type $(i + 1)_{II}$ reading points are equivalent to reading points of type $i_I$ at the previous vertex, as indicated in Fig.(5b). Let us check how these results, gained from one-loop anomaly calculations where all forests were trivial, transfer to the general case. To this end let us study charge conjugation properties.

In a closed fermion loop we always have to sum over both possible charge flows. Using this let us try to prove Furry's theorem. We will use the charge conjugation properties of propagators and vertices as summarized in Eq.(19). As long as one can guarantee a proper behaviour of the axial coupling under charge conjugation one can prove Furry's theorem also in BM-type schemes.

Using Eq.(19) we have expressions of the form

$$\mathrm{Tr}(\gamma_5 (\slashed{l} + \slashed{q}_1)\Gamma_1 \dots \Gamma_k) \text{ clockwise orientation},$$
$$+\mathrm{Tr}(\gamma_5 (\slashed{l} + \slashed{q}_1)\Gamma_k \dots \Gamma_1) \text{ anticlockwise orientation}, \tag{1}$$

where again the $\Gamma_i$ correspond to maximal forest terms, $\slashed{l}$ is the loop momentum of the fermion loop and the $\slashed{q}_i$ are combinations of exterior momenta and other loop momenta induced by the forests.

For the first line we obtain

$$\mathrm{Tr}(\gamma_5 \Gamma_k \dots \Gamma_1 (\slashed{l} + \slashed{q}_1)),$$

which is a cyclic permutation of the second line in Eq.(1). We summarize this result in the following figure:

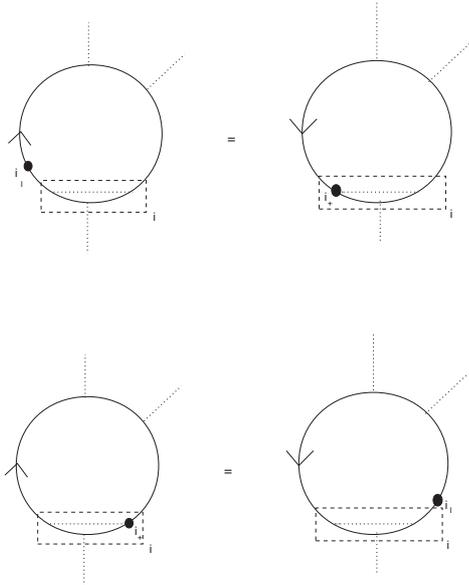

Fig.(7) Furry's theorem.

stating that a $i_I$ reading point with momentum flow clockwise is related to the next reading point in the direction of charge flow and the charge flow then reversed, which is $i_+$ in counterclockwise momentum flow. Similarly one finds that $i_+$ is related to $i_I$ in the reversed orientation.

So we can maintain Furry's theorem by adopting the reading prescription

$$(i_I + i_+)/2. \tag{2}$$

What about type $i_{II}$ reading prescriptions? Clearly we can assign to them a combination with type $j_+$ ($i \neq j$ in general) reading points which maintains Furry's theorem too. Further, Fig.(6) together with Fig.(5a) above seem to show that whenever an exterior vertex $i$ is screened by a subdivergence we can start Trace reading at $i_+$ or $i_I$ without assigning an anomalous behaviour to the $i$-vertex, thus allowing for a much wider class of reading prescriptions. But this is not so. Because whenever a vertex is contained in a maximal forest there is also a counterterm graph replacing this forest where this vertex is unscreened.

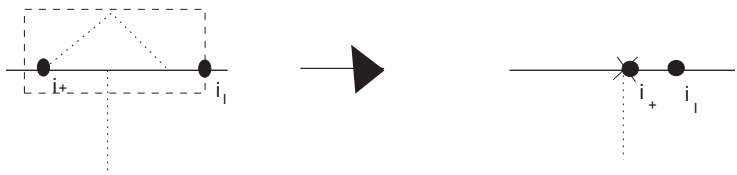

Fig.(8a) Unscreening a vertex by counterterming it.

This also excludes type $i_{II}$ reading points as they are, by the same mechanism, also type $j_I$ for some other vertex $j = i-1$ in the counterterm graphs. So a type $i_{II}$ reading point would assign an anomalous behaviour to the reading point $(i-1)_I$ thus corrupting the renormalization of the vertex $(i-1)$:

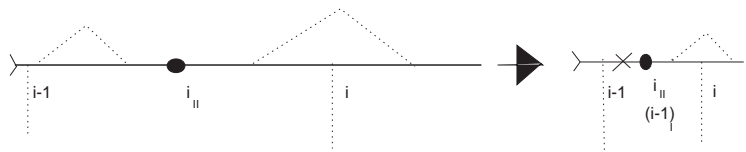

Fig.(8b) The problem with type $i_{II}$ vertices.

To summarize, we have restricted ourself for consistency reasons to reading points of type $i_+, i_I$, where $i$ counts all $A$ vertices. This automatically incorporates vector current conservation and gives the correct results for the anomalies. We have kept Bose symmetry by summing over all these reading points if appropriate and maintained Furry's theorem by using a symmetric reading prescription in the $i_+$ and $i_I$ type reading. So we are left with a single consistent reading prescription, the Bose symmetrized sum of Eq.(2), where the sum runs over all axial couplings to the fermion loop. Note that our reading prescription is based solely on reading points determined by exterior couplings to the fermion loop and is independent from all further topological properties of the actual Feynman graph.

Does this reading prescription necessarily exist? As the determination of reading points uses ZFF we can expect difficulties when we have overlapping divergences for singularities not generated by UV divergences. Consider for example the following two-point function:

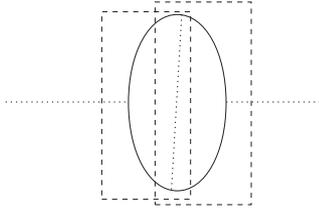

Fig.(9) Overlapping divergences.

As far as its UV singularities are concerned the diagram vanishes and for higher $n$-point functions the overlapping cannot appear. But for this two-point function, appearing as a squared matrix element, it must not vanish (because there is a second momentum involved); however we can guarantee charge conjugation and vector current conservation with our reading prescription, but an anomalous term is likely to appear due to the fact that there is no reading point outside all subdivergences. This is indeed what happens when one considers for example the case of IR divergent phase space integrations in the process $g + W \rightarrow q\bar{q}$ [7].

# 3   BRST Identities

In this section we are concerned with the study of possible violations of BRST identities in our scheme. Such violations would result in modifications of these identities, a situation which is familiar in BM schemes. We claim that such modifications do not appear here.

In the following it is understood that every graph has its subdivergences subtracted, $G \rightarrow \overline{\mathcal{R}}(G)$, as only overall divergences can possibly generate violations of BRST identities in our scheme.

This is so because the $\overline{\mathcal{R}}$ operation is always sufficient to render a $\gamma_5$-odd fermion loop finite in an anomaly free theory, according to our results from the previous section. As all noncyclic results are at least of order $(D-4)$ algebraically, a UV divergence is needed to have possible violations of BRST identities. As UV overall divergences are renormalization scheme independent, we do not have to specify a certain scheme in the following.

Let us start considering the BRST identity for the fermionic vertex. Our notation follows [8].

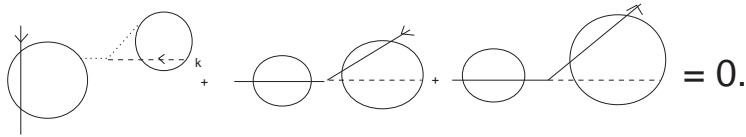

Fig.(10) The BRST identity for the fermionic vertex.

Possible violations of the above identity generated by the $\gamma_5$-problem can only result from a modification of the three-point Green function in such a way that its contraction with the momentum $k$ fails to give the expected two-point fermion propagators involved in the last two terms in the identity. So we will concentrate in the following on the study of this vertex function, assuming that it has a $\gamma_5$-odd fermion loop as a subgraph.

The exterior boson can either couple to the fermion loop or somewhere else.

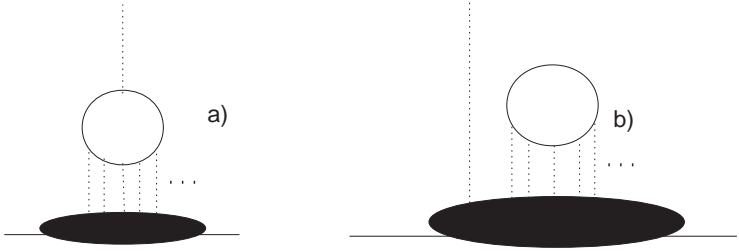

Fig.(11) Radiative corrections to the fermionic vertex. The black oval bubble indicates an arbitrary structure, and the fermion loop under consideration might have arbitrary vertex and self-energy corrections, their subdivergences being subtracted by the $\overline{\mathcal{R}}$ operation already. The same remark applies to the propagators connecting the fermion loop to the bubble.

In BM-type schemes the identity in Fig.(10) is violated because of the non-vanishing anti-commutator. Here, it is not sufficient to show that Fig.(11) fulfills its BRST identities. To this end one would only have to prove that its contraction with the boson momentum results in the corresponding two-point Green functions of the fermionic selfenergy of Fig.(10) above but there would still be a problem at the multiloop level.

To see this consider the following graph as an example:

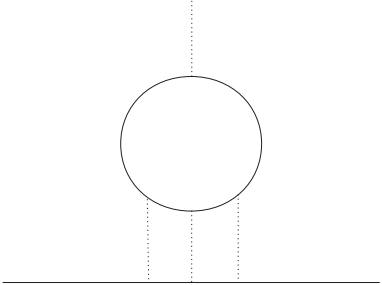

Fig.(12) The generation of a non-anticommuting $\gamma_5$.

After integrating out all loop momenta and evaluating $\mathcal{T}$race the result will be proportional

$$\sim \epsilon_{\nu\mu_1\mu_2\mu_3}\gamma_{\mu_1}\gamma_{\mu_2}\gamma_{\mu_3} \quad \sim \quad \acute{\gamma_5}\gamma_\nu, \;\; \nu \in (0,1,2,3),$$

where $\acute{\gamma_5}$ is a non-anticommuting $\gamma_5$ like in BM schemes.

By naive power counting, the above graph is logarithmic overall divergent. So it might generate a counterterm introducing a non anti-commuting $\acute{\gamma_5}$ and would thus introduce in our scheme all the same problems the BM-schemes have with BRST identities, at least from the three-loop level on. So these counterterms would plague all higher order loop corrections. This would be a serious drawback for our scheme. The only possible cure will be an improved power-counting for these three-point functions.

Note that this does not mean that the fermionic vertex has an improved power counting in general; it is once and for all logarithmic divergent, but what we are looking for is to show the absence of divergences for a certain set of graphs, an improved power counting for graphs containing a closed $\gamma_5$-odd fermion loop as a subgraph.

We do not adopt the standard way of improving power counting by restricting the set of possible Lorentz covariants with the help of gauge invariance. This method is helpful to improve the power counting for a given Green function and could well be applied here. But the sceptical reader might have doubts even if our scheme conserves vector gauge invariance.

Let us study Fig.(11a) first. We have to sum over all possibilities where the exterior boson is connected to the fermion loop. As we are considering a renormalizable theory all UV divergence is located in the form factors $\sim \gamma_\mu$ or $\sim \gamma_5 \gamma_\mu$. To extract the logarithmic overall UV divergence it is sufficient to consider the graph at momentum transfer zero, $q = 0$. This is true because our scheme does not change power-counting. So acting with a first order derivative in an exterior momentum on the graph must kill its logarithmic overall divergence. Then the sum over all couplings appearing in Fig.(13), evaluated with one arbitrarily chosen reading point

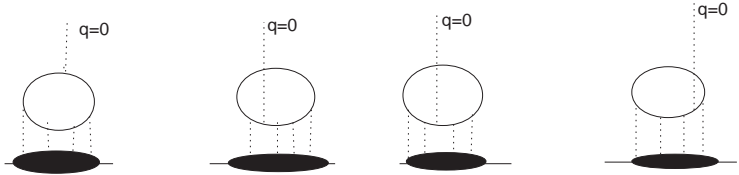

Fig.(13) The exterior boson with vanishing momentum transfer, coupling in all possible ways to the fermion loop.

is a total derivative of Fig.(14)

$$\text{Fig.(13)} \quad \sim \quad \int d^D l \frac{\partial}{\partial l_\rho} \text{Fig.(14)}, \tag{3}$$

and thus vanishes. Note that this argument remains true even if all couplings and propagators in the fermion loop in Fig.(13) are fully dressed ones. The above argument is a generalization of the identity

$$-\frac{\partial}{\partial l_\rho} \frac{1}{\slashed{l} + i\epsilon} = \frac{1}{\slashed{l} + i\epsilon} \gamma_\rho \frac{1}{\slashed{l} + i\epsilon},$$

to our purposes.

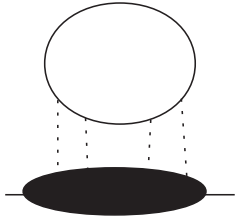

Fig.(14) This graph generates Fig.(13) via a total derivative acting on the integrand.

So we can improve power counting for these graphs from a logarithmic divergent to convergent behaviour and thus see that Fig.(11a) will not give rise to a counterterm.

We cannot use this argument for Fig.(11b), but we will argue in the following way. Consider first the case of a fermion loop with $2k$ couplings in Fig.(11b). It then has an odd number of vector couplings, so the trace over the group generators will be of the form

$$tr(\sum_{\text{perm.}} T_{a_1} \dots T_{a_{2k}}) - tr(\sum_{\text{perm.}} T_{a_{2k}} \dots T_{a_1}), \tag{4}$$

where the sum runs over all permutations generated by the different ways of connecting the fermion loop to the bubble and the two sums correspond to the two possible orientations of charge flow. According to Furry's theorem each summand in one of the above sums appears with the opposite sign in the other sum.

Consider first the case of no abelian factors in the structure group, $SU(N)$ say. The above trace will be purely imaginery (for hermitian generators and imaginery structure constants $f_{abc}$, $[T_a, T_b] = f_{abc}T_c$). For an anomaly safe representation we have $d_{abc} = 0$, and the above expression, being imaginary, must be linear in $f_{abc}$. But we cannot match an even number of indices with products of the $f$ tensor and $\delta_{ab}$, the metric tensor in group space, so it vanishes. In the same way the trace appearing in the case of $2k + 1$ couplings in Fig.(11b) is of the form

$$tr(\sum_{\text{perm.}} T_{a_1}\ldots T_{a_{2k+1}}) + tr(\sum_{\text{perm.}} T_{a_{2k+1}}\ldots T_{a_1}), \qquad (5)$$

so that it is real and cannot contain the $f$ tensor. Again we cannot match an odd number of indices with the $\delta$-tensor alone and again it vanishes.

Now let us generalize to the case of abelian factors in the structure group, e.g. $U(1) \times SU(2)$. The above argument cannot be pursued as an appearance of an odd number of hypercharge generators $Y$ in the trace ruins the above reasoning.

We can refine it in the following manner. Consider Fig.(11b) where we emphasize that it contains the sum over all possible ways of connecting the fermion loop to the graph via the internal bosons. So this sum will be totally symmetric under simultaneous exchange of group indices, Lorentz indices, momenta and boson masses. In the case of $2k$ couplings non-vanishing terms must still be linear in the $f$-tensor, for example $tr(Y^{2k+1}T_aT_bT_c) \sim f_{abc}$ ($d_{abc} = 0$!), so they are necessarily antisymmetric in at least three of the group indices. As these internal indices are to be summed over, and as the overall divergence is independent of the masses of the internal bosons, the result vanishes when summing over the internal group indices. This must not be true for the terms of vanishing overall degree of divergence as for them the antisymmetric part of the trace can become symmetric when multiplying terms in the loop integrals which are possibly antisymmetric under the corresponding exchange of masses, though, as one always can work with mass independent renormalization schemes, this is a rather academic restriction.

In the case of $(2k + 1)$ couplings our trace of the form Eq.(5) will not vanish if, again, we have an odd number of hypercharge generators. But now the result is symmetric. But this means that we only have to consider the symmetric part under simultaneous exchange of Lorentz indices and corresponding momenta. As all outgoing momenta of the fermion loop in Fig.(11b) are to be integrated over with an explicitly symmetric integration measure, we are restricted to the total symmetric part in the exchange of Lorentz indices. But the $\mathcal{T}race$ of the fermion loop is necessarily proportional to the Levi Civita tensor, so again the result vanishes.

We have improved the power counting by at least one degree for Fig.(11a,b) and therefore will not have counterterms generated by these graphs. Further they fulfill their BRST identity because, with a vanishing overall degree of divergence, we can cyclically permute $\mathcal{T}race$ without changing the results, thus directly establishing the BRST identity in Fig.(10). Note that the above reasoning is not restricted to the case $U(1) \times SU(2)$ or $U(1) \times SU(2) \times SU(3)$. It applies to arbitrary abelian factors as long as we have an anomaly free representation.

Let us check the other BRST identities.

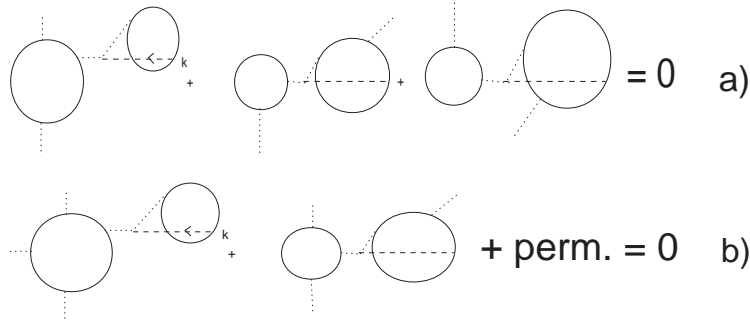

Fig.(15) The other BRST identities for vertices [8].

For the logarithmic divergent functions Fig.(15b) we can use the above mechanism Eq.(3) to reduce the problem to the case of no exterior particle coupling to the fermion loop. The arguments we used for Fig.(11a) can be used almost unchanged for the case of several exterior lines coupling to the fermion loop. The novel feature that there is more than one exterior line at the fermion loop does not demand a separate treatment. We are still allowed to set one of the momentum transfers to zero and could then consider appropriate sums over all possible couplings. Here we used the fact that all possibilities to connect $j$ exterior particles to the fermion loop can be ordered in the following manner. First connect $(j-1)$ lines in some way to the fermion loop. Then connect the last line (with momentum transfer zero) in all possible ways and apply the above total derivative argument Eq.(3) to this sum. Now sum over all possible ways to connect the $(j-1)$ lines.

We are left with the case that no exterior line couples to the fermion loop, so that there is a total internal fermion loop. For this case we then use the arguments we used for Fig.(11b) as they are totally independent of the nature of the exterior particles. Let us call this the internal fermion loop argument (IFLA).

In the case of linear divergences we can use the same arguments as before. Though there is a linear divergence the relevant form factor for the overall degree of divergence always factorizes an exterior momentum. We can still achieve the graph to be finite by acting with an appropriate first order derivative in an exterior momentum on it. So we can apply the total derivative argument Eq.(3) to the case of exterior couplings to the fermion loop and apply again IFLA for the remaining case.

The last identity to be checked is the BRST identity for the two-point gauge boson propagator. We are confronted with a quadratic divergence. We cannot apply the total derivative argument.

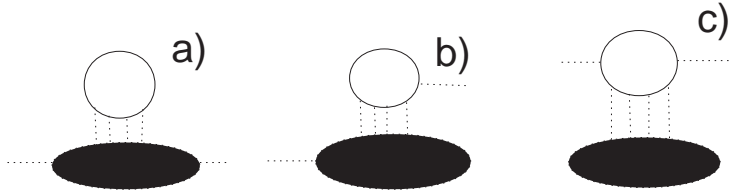

Fig.(16) The three cases for the gauge boson propagator.

But nevertheless, for the case of one or two external lines coupling to the fermion loop Fig.(16b, 16c) the result follows easily. The only non-trivial case is when there are terms proportional to $\Gamma^3_{5,\mu}$ or $\Gamma^5_{5,\mu}$ in Fig.(16b) resp. $\Gamma^4_5$ in Fig.(16c). Here $\Gamma^3_{5,\mu}, \Gamma^5_{5,\mu}, \Gamma^4_5$ are short-hand notations for a contracted product of the Levi-Civita tensor with 3, 5 or 4 $\gamma$-matrices as defined in the appendix. These structures can be generated by the black bubble in Fig.(16). Then we can explicitly verify transversality by considering $\mathcal{T}r$aces of the form

$$\mathcal{T}r(\gamma_5\gamma_\mu \slashed{l}\Gamma^4_5(\slashed{l}+\slashed{q})), \ \ \mathcal{T}r(\gamma_5\gamma_\mu \slashed{l}\Gamma^5_{5,\mu}(\slashed{l}+\slashed{q})), \ \ \mathcal{T}r(\gamma_5\gamma_\mu \slashed{l}\Gamma^3_{5,\mu}(\slashed{l}+\slashed{q})),$$

which all vanish identically when contracted with $q^\mu$.

For Fig.(16a) we use IFLA again.

We conclude that we are not confronted with modifications of BRST identities in our scheme. Also uniqueness of the $\mathcal{T}r$ace functional follows easily. The above reasoning covers all relevant couplings of exterior particles to fermion loops. Cases like the Yukawa coupling in the Higgs-fermion sector are unproblematic as for the overall divergent Green functions (see [8] section 4.2.5) the above arguments can be applied in a similar manner. Other Green functions involve particles not coupling to fermions and can most easily handled by the IFLA argument.

# 4 Verification of the Adler Bardeen Theorem in this Scheme

Here we will give a short proof of the Adler Bardeen Theorem [9]. We can restrict our attention to the case of radiative corrections to the axial vector coupling. These can be generated by overall divergences of the corresponding graphs or as subdivergences modifying the axial vector coupling to the fermion loop.

We first note that whenever the $A$ vertex is located in the interior of a non-trivial maximal forest it is screened by this forest. This means that a contraction with the corresponding momenta establishes the corresponding BRST identity, as our reading rules guarantee that we do not start $\mathcal{T}race$ reading inside this forest. On the other hand we still have vector current conservation, so we will not find an anomaly in these higher loop graphs.

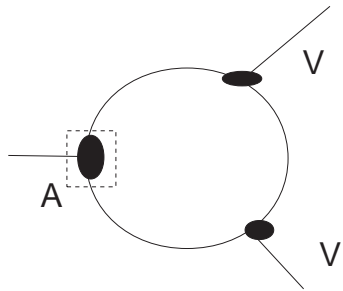

Fig.(17) The screening. Our type $i_+$ or $i_l$ reading points give a non-anomalous result.

In case that the maximal forest containing $A$ contains the whole fermion loop we can always find a reading point guaranteeing a non-anomalous result:

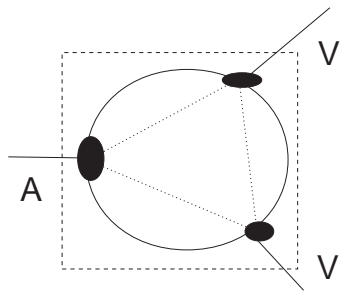

Fig.(18) The maximal forest contains the fermion loop. Any reading point which does not coincide with an exterior vertex or an incoming propagator of such a vertex is anomaly free.

It remains to discuss the case that the fermion loop is part of a bigger topology, Fig.(19).

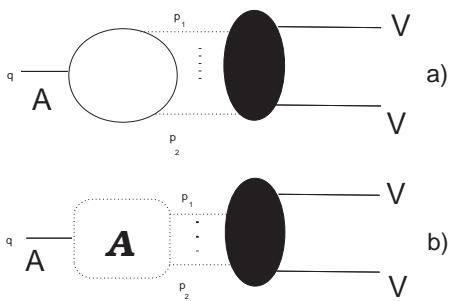

Fig.(19) a) The fermion loop appearing as a subgraph in an anomalous Green function.
b) Integrating out the fermion loop gives the anomaly or a related tensor.

We cannot use the above argument, as there can be overall divergences which, in contrast to subdivergences of the fermion loop, do not shelter the $A$ vertex from becoming anomalous.

At least one fermion loop must necessarily appear as a subgraph involving an odd number of $A$ vertices to generate a possible anomalous structure. Then this graph gives us the anomaly as a result, see Fig.(19a). The case that it couples to more than three vertices -the dots in Fig.(19a)- does not modify the following reasoning in an essential manner. The result for this graph will still involve the Levi Civita-tensor, and thus sufficient antisymmetry for the following argument to apply:

Integrating out the fermion loop, the resulting expression, Fig.(19b), is only logarithmic divergent by power counting, as the anomaly behaves as $\epsilon^{0123}_{\mu_1\mu_2\beta_1\beta_2} p_1^{\beta_1} p_2^{\beta_2}$ where $p_1, p_2$ are the momenta entering the other ($V$) vertices of the fermion loop. But we can always choose these momenta which contract the $\epsilon$-tensor to have the form $p_1 = k + q_1, p_2 = k + q_2$, that is they are both combination of the same internal loop momentum $k$ and different exterior momenta $q_i$. We see that by contracting with the Levi-Civita tensor we improve power counting by one, so that our graphs have vanishing overall degree of divergence. They then vanish because the $Trace$ evaluations bring at least an order $(D-4)$ operator. We conclude that there are no radiative corrections to the one-loop anomalous fermion loops - the Adler Bardeen theorem.

# 5 General Remarks and Conclusions

In this paper we examined some basic features of a new proposal for the $\gamma_5$-problem in DR. Compared to the common approach of a non-anticommuting $\gamma_5$ we find a different behaviour with respect to gauge symmetries. The anomalies appear as the only consequence of the breaking of Lorentz invariance in this new scheme. All further modifications of the BRST identities could be avoided. In this sense, this new scheme seems most economical in its treatment of gauge invariance.

This has to be compared with BM schemes. There even in amplitudes involving only open fermion lines one has a modification of the BRST identities. Modifications in this sector are not possible in our scheme. Now our reasoning in section 3 was based on an improvement of power-counting. So the arguments there should have consequences for BM schemes too. In the open fermion line sector there can be no doubt that the BM schemes, after an appropriate modification of their BRST identities, will give the same results for physical amplitudes as our scheme gives. The modifications cure the spurious anomalies of the gauge symmetry in the BM scheme. But then all our reasoning in section 3 should be applicable to the BM schemes too, showing that in processes involving closed fermion loops the modified BM BRST identities are sufficient to incorporate all effects of the $\gamma_5$-problem. This seems reasonable because all subdivergences of a fermion loop are open fermion line amplitudes. Indeed, for example the conclusions in [10] or [11] seem to support this viewpoint. It was found that, after compensating the open fermion line spuriosities of BM schemes in an appropriate manner, it is justified to calculate with an anticommuting $\gamma_5$ in closed fermion loops in the case of several $\gamma_5$'s [10]. In [11] this property was checked and confirmed to all loop orders, as expected from the results in [4].

All this is certainly no surprise from the viewpoint of our scheme, and might be considered as justified by our reasoning in section 3.

Note that one cannot derive such results from schemes favouring an anticommuting $\gamma_5$ naively [12].

It should be justified then to consider BM schemes and the new scheme as fully equivalent. It seems that from the viewpoint of the practitioner the new scheme is more efficient [6]. The only complication is the need to determine all maximal forests. On the other hand this is an unavoidable step in the calculation of multiloop diagrams anyhow, so there is no real complication in the $D$-dimensional generalization of the $\gamma$-algebra. Especially statements of the form that chiral fermions cannot be consistently defined in DR are not justified.

When one applies DR as a regulator to nonrenormalizable theories the considerations of section 3 certainly do not apply. To use DR as a unique regulator demands some physical input then. An example is the demand of vector current conservation for the one-loop triangle anomaly. Such physical input is necessary in all schemes, and is incorporated in the BM-scheme explitly (do not commute $\gamma_5$ away from the axial vertex) as well as in our scheme (start $\mathcal{T}$race reading at the axial vertex).

Also IR regularizations do not obey the arguments of section 3, but all ambiguity can be avoided by treating squared matrix elements and corresponding loop amplitudes in the same manner [6]. This does not exclude IR anomalies as mentioned at the end of section 2.

Two final remarks. The first considers the rôle of the Fierz transformation. The use of it in BM schemes is clearly demonstrated in [13]. To make use of the formalism obtained there in our scheme we stress that the derivation of Fierz identities between bilinear scalar spinor densities is not restricted to ordinary traces appearing as the linear functional in them. A short glance to the standard derivation shows that one can generalize to arbitrary linear functionals. Especially we can choose this linear functional to be $\mathcal{T}$race to obtain

$$\mathcal{T}r(\overline{\psi_2} \otimes \Gamma^i(\psi_1))\mathcal{T}r(\overline{\psi_4} \otimes \Gamma_i(\psi_3)) = \sum_j c_{ij}(D)\mathcal{T}r(\overline{\psi_2} \otimes \Gamma^i(\psi_3))\mathcal{T}r(\overline{\psi_4} \otimes \Gamma_j(\psi_1)). \qquad (6)$$

The matrix $c_{ij}(D)$ is the $D$-dimensional generalization of the matrix appearing in the familiar Fierz transformations in four dimensions [13]. Then one can make use of the formalism in [13] to obtain a unique prescription for the handling of Fierz identities in our scheme. Due to the projector properties of $\mathcal{T}$race the sum in Eq.(6) will become a finite sum if the involved spinors are known to be on-shell.

The final remark provides an example how to summarize the gauge properties of the new scheme in a kind of toy homology. In the following we neglect the ghost sector as we do expect all anomalous contributions to be generated by the treatment of a $\gamma_5$-odd fermion loop.

Assume we act with the BRST identity Fig.(10) on an arbitrary Green function $G^{(k)}$ involving a fermion loop with $k$ exterior couplings to this fermion loop. We can then establish a relation between this Green function and two Green functions $G^{(k-1)}$, with only $(k-1)$ couplings to the fermion loop, as usual

$$q_i^{\mu_i} G^{(k)}_{\mu_1 \ldots \mu_i \ldots \mu_k}(q_1, \ldots q_k) - G^{(k-1)}_{\mu_1 \ldots \hat{\mu_i} \ldots \mu_k}(\{1\}) + G^{(k-1)}_{\mu_1 \ldots \hat{\mu_i} \ldots \mu_k}(\{2\}) = 0, \qquad (7)$$

where $\{1\}, \{2\}$ specify appropriate sets of exterior momenta. Define now an operator $\Delta := \sum_i \Delta_i$, acting on fermion loops coupling to exterior lines, where the sum runs over all (axial) couplings of exterior lines to the fermion loop. We define

$$\Delta_i := q_i^{\mu_i} G^{(k)}_{\mu_1 \ldots \mu_i \ldots \mu_k}(q_1, \ldots q_k) - G^{(k-1)}_{\mu_1 \ldots \hat{\mu_i} \ldots \mu_k}(\{1\}) + G^{(k-1)}_{\mu_1 \ldots \hat{\mu_i} \ldots \mu_k}(\{2\}). \qquad (8)$$

It results that every fermion loop Green function $G^{(k)}$ which fulfills its BRST identities is in the kernel of this operator, $\Delta(G^{(k)}) = 0$. If we have a Green function $G^{(0)}$ for which no exterior line is connected to a closed fermion loop, IFLA justifies the definition $\Delta(G^{(0)}) = 0$.

Further we have $\Delta^2 = 0$ in our $\gamma_5$-scheme. This is so because one can cut a closed fermion loop only at one point. The operator $\Delta^2$ corresponds to two contractions of a fermion loop with exterior momenta. For such a contraction to fail in a resulting BRST identity the fermion loop must be cut open at the corresponding exterior vertex, cf. Fig.(6). So at least one of the two contractions will fulfill its BRST identity, so that $\Delta^2 = 0$ in our $\gamma_5$-scheme.

Consider anomalies $< A >$. They are generated by $\Delta$, by construction of $\Delta$. So they are in the image of $\Delta$, for example the AAA anomaly is $\Delta(G^{(3)}_{\text{oneloop}})$. So the anomalies are in the image modulo the kernel of $\Delta$, $< A >\in H_\Delta \equiv Im_\Delta/Ker_\Delta$. This is just the statement that the addition of BRST invariant terms leaves the anomalies unchanged. Though the above homology is a very naive and truncated counterpart of the usual BRST cohomology which specifies physical amplitudes it summarizes the content of our $\gamma_5$-scheme in a neat manner.

One can also write down a path-integral representation for the operator $\Delta$ based on the cyclic permutation group, Wick's theorem and the generating functional for closed fermion loops. If anything of interest, like connections to cyclic (co)homology, can be obtained from it will be a subject of future studies.

Acknowledgements
Stimulating discussions on the subject with D. Broadhurst, A. Buras, J. G. Körner, K. Schilcher and G. Thompson are gratefully acknowledged. It is a pleasure to thank P. D. Jarvis especially for discussions on the (co)homological properties of the BRST identities.

But most grateful I am to R. Delbourgo. His insight and comments were of enormous help.

I would like to thank D. Maison and C. Schubert for correspondence.

T. Baker, N. Jones and T. Waites were helpful in generating all the figures.

This work was supported under grant number A69231484 from the Australian Research Council.

# A    Definition of the Scheme

Consider an arbitrary string of $\gamma$-matrices which has a natural $\mathbf{Z_2}$ grading

$$
\begin{aligned}
\Gamma_0 &\equiv \mathbf{1}, \\
\Gamma_{2k} &\equiv \gamma_{\mu_1}\ldots\gamma_{\mu_{2k}}, \\
\Gamma_{2k+1} &\equiv \gamma_{\mu_1}\ldots\gamma_{\mu_{2k+1}}.
\end{aligned}
\tag{9}
$$

We can expand such a string according to a basis including the identity as a zero form and the antisymmetric products of $\gamma$-matrices.

$$
\begin{aligned}
\Gamma_{2k} &\equiv a_{2k}^0\mathbf{1} + a_{2k}^2\gamma\wedge\gamma + a_{2k}^4\gamma\wedge\gamma\wedge\gamma\wedge\gamma + \ldots \\
\Gamma_{2k+1} &\equiv a_{2k}^1\gamma + a_{2k}^3\gamma\wedge\gamma\wedge\gamma + \ldots.
\end{aligned}
\tag{10}
$$

In four dimensions the expansion stops with the four-form $\gamma\wedge\gamma\wedge\gamma\wedge\gamma = \gamma_0\gamma_1\gamma_2\gamma_3$. Note that we normalized our forms $\omega \equiv \gamma, \gamma\wedge\gamma, \ldots$ so that $\omega^2 = \mathbf{1}$.

Now we have in *four* dimensions with Eq.(10)

$$
\begin{aligned}
tr(\Gamma_{2k+1}) &= 0, \\
tr(\Gamma_{2k}) &= a_{2k}^0 tr(\mathbf{1}), \\
tr(\gamma_5\Gamma_{2k+1}) &= 0, \\
tr(\gamma_5) &= 0, \\
tr(\gamma_5\Gamma_2) &= 0, \\
tr(\gamma_5\Gamma_4) &= a_4^4 tr(\mathbf{1}), \\
tr(\gamma_5\Gamma_{2k}) &= a_{2k}^4 tr(\mathbf{1}) \text{ for } k > 2, \\
\text{where} \quad & \text{i}\gamma_5 \ \gamma_0\gamma_1\gamma_2\gamma_3 = \mathbf{1}, \\
a_4^4 &= \epsilon_{\mu_1\ldots\mu_4}^{0123}, \text{ and } a_{2k}^4 \text{ is a linear combination} \\
& \text{of the } \epsilon\text{-tensor and metrical tensors:} \\
a_{2k}^4 &= \epsilon_{\alpha_1\ldots\alpha_4}^{0123} tr(\gamma_{\alpha_1}\ldots\gamma_{\alpha_4}\gamma_{\mu_1}\ldots\gamma_{\mu_{2k}}).
\end{aligned}
\tag{11}
$$

The crucial point is that in this way we can express the trace operation as a projection, which projects on the $a_{2k}^0$ element if there is an even number of $\gamma_5$ in the argument and on the $a_{2k}^4$ element if the number of $\gamma_5$ is odd.

Let us consider the generalization to $D \neq 4$ dimensions. In a slight abuse of notation $\gamma_5$ still denotes the one unique anticommuting element of the Clifford algebra:

$$\{\gamma_5, \gamma_\mu\} = 0.$$

Note that with this $\gamma_5$ we would have

$$tr(\gamma_5 \gamma_{\mu_1} \ldots \gamma_{\mu_{2k}}) = 0 \ \forall \ 2k < D \text{ and } D \text{ integer}.$$

An explicit representation of dimensional regularization can be given in terms of functions which depend on the dimensionality of spacetime in a manner which clarifies the continuation to arbitrary complex dimensions; this is done by formulating DR on an arbitrary dimensional vector space [14]. We stress that such a consistent continuation is always based on Clifford algebras in higher integer dimensional spacetimes [14].

Having this in mind we see that the only modification we get if the expansion in Eq.(10) does not stop with four-forms in four dimensions is that our linear projection operation is not a trace anymore, but a linear functional which happens to agree with the trace functional in four dimensions. So this functional, which we denote by $\mathcal{T}\!r$ in the following, acts as

$$\mathcal{T}\!r(\Gamma_{2k}) = a_{2k}^0 \mathcal{T}\!r(\mathbf{1}), \tag{12}$$

$$\mathcal{T}\!r(\gamma_5 \Gamma_{2k}) = a_{2k}^4 \mathcal{T}\!r(\mathbf{1}). \tag{13}$$

Note that with an anticommuting $\gamma_5$ we have to consider these two cases only. The existence of such an anticommuting element is guaranteed by the periodicity properties of Clifford algebras [15]. Note that the notation $\epsilon^{0123}$ emphasizes that our projection is on the four dimensional Minkowski space represented by the index set $(0, 1, 2, 3)$ (there are other four dimensional hyperplanes in a higher dimensional vector space and so there are other four forms as well as other evanescent forms of higher degree). That means the Levi-Civita tensor we use is the Levi-Civita of four dimensional space time imbedded in the higher dimensional vectorspaces so that it vanishes whenever one of its indices is not in the index set $(0, 1, 2, 3)$. This can be summarized in the definition

$$\epsilon^{0123}_{\mu_1 \mu_2 \mu_3 \mu_4} \equiv \delta^0_{[\mu_1} \delta^1_{\mu_2} \delta^2_{\mu_3} \delta^3_{\mu_4]}, \tag{14}$$

which expresses the Levi-Civita as an antisymmetrized product of Kronecker $\delta$'s.

Note further that the expansions in Eq.(10) gives the coefficients $a_{2k}^4$ as products of metrical tensors times $\epsilon^{0123}$. No (evanescent) forms of rank $> 4$ appear. So these coefficient functions remain unchanged for fixed $k$ when we vary the spacetime dimensionality in the sense of [14]. This means that our functional $\mathcal{T}\!r$ can be defined in a consistent manner in DR.

Let us mention here a convenient way to calculate $\mathcal{T}\!r$. By its definition and the $\mathbf{Z}_2$-grading of the Clifford algebra we have

$$
\begin{aligned}
\mathcal{T}\!r(\Gamma_{2k+1}) &= 0, \\
\mathcal{T}\!r(\gamma_5 \Gamma_{2k+1}) &= 0, \\
\mathcal{T}\!r(\gamma_5) &= 0, \\
\mathcal{T}\!r(\gamma_5 \Gamma_2) &= 0, \\
\mathcal{T}\!r(\Gamma_{2k}) &= tr(\Gamma_{2k}),
\end{aligned} \tag{15}
$$

so that $\mathcal{T}\!r$ is cyclic when acting on strings containing an even number of $\gamma$-matrices and no $\gamma_5$. Further with Eq.(10)

$$
\begin{aligned}
\epsilon^{0123}_{\mu_1 \ldots \mu_4} \gamma_{\mu_1} \ldots \gamma_{\mu_4} \Gamma_{2k} &= [\epsilon^{0123}_{\mu_1 \ldots \mu_4} \gamma_{\mu_1} \ldots \gamma_{\mu_4}] a_{2k}^0 \mathbf{1} + \\
&\quad [\epsilon^{0123}_{\mu_1 \ldots \mu_4} \gamma_{\mu_1} \ldots \gamma_{\mu_4}] a_{2k}^2 \ \gamma \wedge \gamma + \\
&\quad a_{2k}^4 \mathbf{1} + \\
&\quad \text{evanescent terms.}
\end{aligned} \tag{16}
$$

Applying $\mathcal{T}r$ to this equation yields

$$\epsilon^{0123}_{\mu_1\ldots\mu_4}\mathcal{T}r(\gamma_{\mu_1}\ldots\gamma_{\mu_4}\Gamma_{2k}) = a^4_{2k}\mathcal{T}r(\mathbf{1}), \tag{17}$$

and using the above Eq.(15) and the definition of $\mathcal{T}race$,

$$\begin{aligned}
\mathcal{T}r(\gamma_5\Gamma_{2k}) &= a^4_{2k}\mathcal{T}r(\mathbf{1}) = \epsilon^{0123}_{\mu_1\ldots\mu_4}\mathcal{T}r(\gamma_{\mu_1}\ldots\gamma_{\mu_4}\Gamma_{2k}) \\
&= \epsilon^{0123}_{\mu_1\ldots\mu_4}tr(\gamma_{\mu_1}\ldots\gamma_{\mu_4}\Gamma_{2k}),
\end{aligned} \tag{18}$$

which gives an easy rule to calculate $\mathcal{T}r$ in terms of the ordinary trace. Note that this does not mean that $\mathcal{T}r$ becomes cyclic or that we are calculating with an not-anticommuting $\gamma_5$ = $i\gamma_0\gamma_1\gamma_2\gamma_3$, it only states that we can use $\gamma_0\gamma_1\gamma_2\gamma_3$ to mimic the projection properties of $\mathcal{T}r$ after using the anticommutation relation $\{\gamma_5,\gamma_\mu\} = 0$ to bring $\gamma_5$ to the left.

One can now easily check that $\mathcal{T}r$ is a noncyclic functional

$$\mathcal{T}r(\Gamma_j\gamma_5\Gamma_i) \neq \mathcal{T}r(\gamma_5\Gamma_i\Gamma_j), \quad (i+j) \text{ even}.$$

An example might be in order here. As we see the first possible non-cyclicity can appear when applying $\mathcal{T}r$ to a string containing an odd number of $\gamma_5$ and six $\gamma$-matrices. The resulting noncyclicity might be regarded as the source of the anomaly in this scheme [16, 4, 6]:

$$\begin{aligned}
\mathcal{T}r(\gamma_5\gamma_{\mu_1}\ldots\gamma_{\mu_6}) &- \mathcal{T}r(\gamma_{\mu_6}\gamma_5\gamma_{\mu_1}\ldots\gamma_{\mu_5})g^{\mu_1\mu_6} \\
&= 8(D-4)\epsilon^{0123}_{\mu_2\cdots\mu_5}.
\end{aligned}$$

Let us mention a few properties of explicit noncyclic $\mathcal{T}races$. The origin of the noncyclicity is the $D$-dimensional metric tensor in the sense that iff the index which is permuted is to be contracted with an index which is restricted to the set $(0,1,2,3)$ all noncyclic effects disappear. The permuted index has necessarily to be contracted to make an effect and after integrating out all loop momenta the only tensor which provides indices not in the Minkowski set is the metric tensor. The Levi Civita tensor we have defined to be nonvanishing only in this Minkowski set while all exterior momenta are restricted to this set in DR [14]. This automatically means that all non-cyclic effects are of $\mathcal{O}(D-4)$ and are restricted to contractions with metric tensors involving a permuted index.

Finally, let us mention some charge conjugation properties. We have

$$\begin{aligned}
\mathcal{T}r(X) &= \mathcal{T}r(CXC^{-1})\forall \text{ strings } X, \\
\mathcal{T}r(X) &= \mathcal{T}r(X^T)\forall \text{ strings } X, \\
C\gamma_\mu C^{-1} &= -\gamma_\mu^T, \\
C\gamma_5 C^{-1} &= \gamma_5{}^T,
\end{aligned} \tag{19}$$

where the last property for $\gamma_5$ holds in $4, 8, \ldots$ dimensions and has to be replaced by $C\gamma_5 C^{-1} = -\gamma_5{}^T$ in $2, 6, \ldots$ dimensions.

For later use we define

$$\begin{aligned}
\Gamma^4_5 &:= \epsilon^{0123}_{\mu_1\mu_2\mu_3\mu_4}\gamma^{\mu_1}\gamma^{\mu_2}\gamma^{\mu_3}\gamma^{\mu_4}, \\
\Gamma^3_{5,\mu} &:= \epsilon^{0123}_{\mu\mu_2\mu_3\mu_4}\gamma^{\mu_2}\gamma^{\mu_3}\gamma^{\mu_4}, \\
\Gamma^5_{5,\mu} &:= \epsilon^{0123}_{\mu_1\mu_2\mu_3\mu_4}\gamma_\mu\gamma^{\mu_1}\gamma^{\mu_2}\gamma^{\mu_3}\gamma^{\mu_4}.
\end{aligned} \tag{20}$$

# B  Zimmermann's Forest Formula

For the renormalization procedure, we need a cancellation mechanism for one-particle irreducible (1PI) subdivergences. So we recall the key properties of Zimmermann's *forest formula* (ZFF) [17].

The renormalization problem is mainly to establish a recursive definition of a cancellation mechanism which guarantees all divergences and subdivergences are absorbed in the redefinition of the coefficients of the monomials of the bare lagrangian. This mechanism is shown to exist with zero-momentum subtractions in [18], with the help of DR together with minimal subtraction in [1, 19], and, in a regularization independent manner (on integrands), in [20].

The above recursion problem has a solution found in [17], where a subtraction procedure is given by application of a subtraction operator to all possible forests of a graph. This is, especially for the problem of overlapping divergences, an elegant method, very powerful for developing proofs with respect to the substructure of graphs.

Let a graph $G$ be given. Let $\mathcal{R}(G)$ denote its regularized value, $\mathcal{U}(G)$ the unregularized graph and let $\mathcal{T}$ denote a mechanism which extracts the divergences (poles) of the graph: $\mathcal{R}(G) = \mathcal{U}(G) - \mathcal{T} \circ \mathcal{U}(G)$ (our notation strictly follows [14]). Let us further assume that we know how to extract subdivergences. Let the graph with subtracted subdivergences be $\overline{\mathcal{R}(G)}$. Then we can define an overall counterterm $\mathcal{C}(G) := -\mathcal{T} \circ \overline{\mathcal{R}(G)}$. We define $\mathcal{C}(G) = 0$ if there is no overall divergence. We have

$$\mathcal{R}(G) = \overline{\mathcal{R}(G)} + \mathcal{C}(G). \tag{21}$$

Now we come to the definition of $\overline{\mathcal{R}(G)}$. By definition, if $G$ has no subdivergences at all, we must have $\overline{\mathcal{R}(G)} = \mathcal{U}(G)$. If it has subdivergences, it is given by

$$\begin{aligned} \overline{\mathcal{R}(G)} &= \mathcal{U}(G) + \sum_{\gamma \subset G} \mathcal{C}_\gamma(G) \\ \mathcal{C}(\gamma) &= \left\{ \begin{array}{l} -\mathcal{T} \circ \overline{\mathcal{R}(\gamma)}, \text{ if } \gamma \text{ has an overall divergence} \\ 0 \quad, \text{ if } \gamma \text{ has no overall divergence} \end{array} \right\}, \end{aligned} \tag{22}$$

where $\gamma$ denotes real subgraphs of $G$. This recursion ends with the one-loop graphs which do not contain further subdivergences. This is the subtraction mechanism which every regularization prescription has to respect.

Now let us turn to the forest formula. Imagine you denote each application of the subtraction operator $\mathcal{T}$ in the above recursion graphically by encircling the corresponding 1PI-subgraph with a box. Call this a forest. The graph $G$ itself without subtraction corresponds to the empty set. Then the set of all forests of a graph includes so-called normal forests, which subtract subdivergences, and so-called full forests, which combine to subtract the overall divergences. Full forests always encircle the whole of $G$. Maximal forests are normal forests not contained in any other normal forest.

Let $\mathcal{F}(G)$ be the set of all forests. Note that with the above definition forests are always disjoint or nested, but not overlapping. Then the *forest formula* is given by

$$\mathcal{R}(G) = \sum_{U \in \mathcal{F}(G)} \prod_{\gamma \in U} (-\mathcal{T}_\gamma) G. \tag{23}$$

The application of this formula gives the same result as the above recursion prescription [17].

It can be proven that these methods for subtracting subdivergences give local counterterms. That is the counterterms are polynomial expressions in exterior momenta of degree not higher than the overall degree of divergence which one obtains from power-counting. Further the counterterms add to a counterterm lagrangian including only the same monomials as in the bare lagrangian, as promised.

## C  Reading prescriptions

We denote the point where we start $\mathcal{T}$race reading as a reading point and any combination of the form $\frac{1}{n}$ times $n$ reading points as a reading prescription.

Consider the $i$-th exterior vertex of a $n$-point fermion loop and the associated maximal forest.

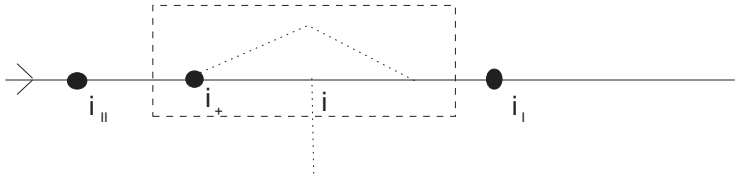

Fig.(C1) Reading points

The non-forbidden reading points are $i_+, i_I, i_{II}$. These are the ones which can be defined with respect to the exterior vertex $i$ as the points which are the first vertex of the maximal forest of $i$, $-i_+-$, the propagator entering this maximal forest, $-i_{II}-$, and the propagator leaving it, $-i_I$. If there are no subdivergences at the exterior vertex $i$ we associate it with the empty forest, so that $i_+$ is the vertex $i$ itself for example. These reading points are considered as to be outside the maximal forest. Any other reading point between $i_+$ and $i_I$ we define to be inside the maximal forest. Such reading points cut open the subdivergence associated with the maximal forest.

Reading points which cannot be defined with respect to an exterior vertex $i$ are called type $III$ reading points. They are useless as they do not provide a unique way of identifying them. To a given loop order the number of $i_+$ and $i_I, i_{II}$ reading points is simply $k$, the number of exterior couplings to the fermion loop, while the number of type $III$ reading points is not fixed at all, though bounded by the loop order. If for some reason we were forced to make use of type $III$ reading points we would have to proof that all there ambiguities do cancel. Fortunately we are able to avoid their use and can omit this cumbersome proof.

Reading points associated with exterior scalar or pseudoscalar vertices are denoted as type III too, stressing that they also can be avoided. They might be dangerous as they do not provide us with a free Lorentz index to identify them. As overall divergences tend to be independent of external momenta the unique identification of such a vertex in a fermion loop is critical.

The reader might wonder where one should start $Trace$ reading then in a $\gamma_5$-odd fermion loop which has no exterior $A$ vertices but only some pseudoscalar vertices. But note that for a non-cyclic effect to appear we need at least six $\gamma$-matrices. As the replacement of axial-vector couplings by pseudo-scalar ones lowers the number of $\gamma$-matrices in the fermion loop we do not have an overall degree of divergence for such a fermion loop itself. So all reading points outside the forests are equivalent. On the other hand, the appearance of such a loop as a subgraph in other loops is covered by our considerations in section 3 (IFLA, total derivative argument).